\documentclass[sigconf, authorversion]{acmart}
\AtBeginDocument{%
  \providecommand\BibTeX{{%
    \normalfont B\kern-0.5em{\scshape i\kern-0.25em b}\kern-0.8em\TeX}}}
    
\begin{document}
\title{AugLimb: Compact Robotic Limbs for Human Augmentation}

\author{Zeyu Ding$^1$\hspace{1em} Shogo Yoshida$^1$\hspace{1em} Toby Chong$^2$ \\ \hspace{1em} Tsukasa Fukusato$^2$\hspace{1em} Takuma Torii$^1$\hspace{1em} Haoran Xie$^1$ }
\affiliation{%
  \institution{Japan Advanced Institute of Science and Technology$^1$\\ The University of Tokyo$^2$}
  \country{Japan}
}
\email{xie@jaist.ac.jp}
\renewcommand{\shortauthors}{Z.Ding, et al.}

\begin{abstract}
 This work proposes a compact robotic limb, AugLimb\footnote{project homepage: https://bit.ly/3mLYcgd}, that can augment our body functions and support the daily activities. AugLimb adopts the double-layer scissor unit for the extendable mechanism which can achieve 2.5 times longer than the forearm length. The proposed device can be mounted on the user's upper arm, and transform into compact state without obstruction to wearers. The proposed device is lightweight with low burden exerted on the wearer. We developed the prototype of AugLimb to demonstrate the proposed mechanisms. We believe that the design methodology of AugLimb can facilitate human augmentation research for practical use.
\end{abstract}





\begin{teaserfigure}
  \includegraphics[width=\textwidth]{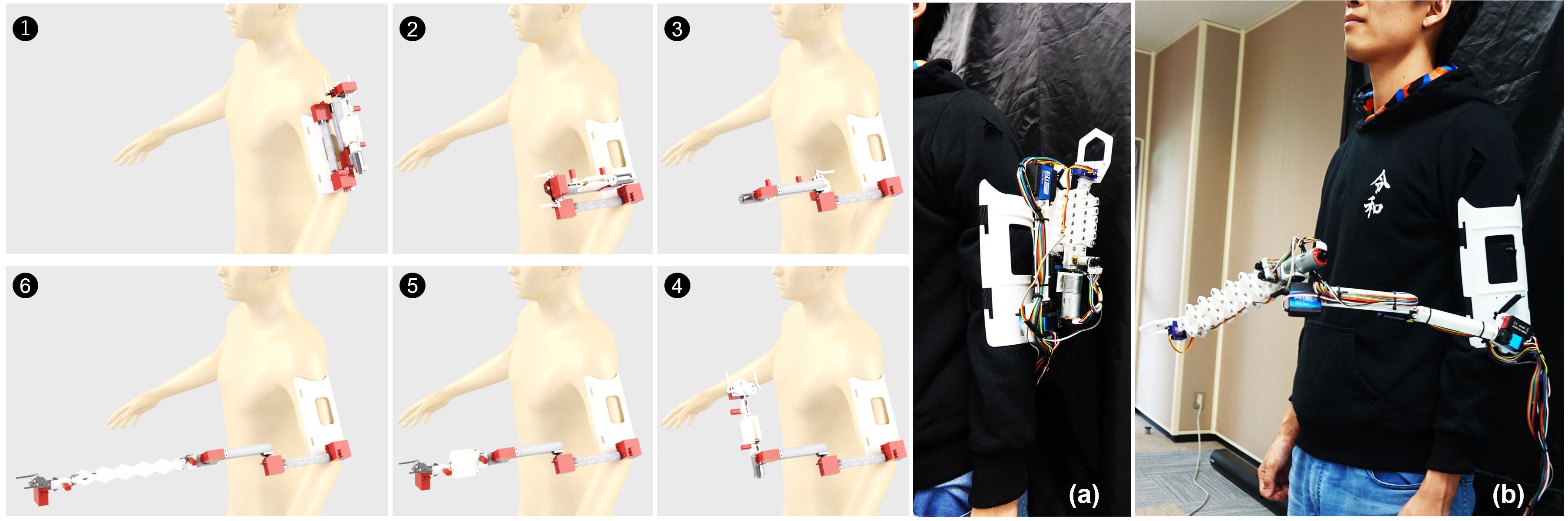}
  \caption{AugLimb robotic arm can transform from a compact size to an extended length from step 1 to 6. We confirmed the usage of the developed prototype mounted on user's upper arm in both compact (a) and expanded states (b).}
  \label{fig:teaser}
\end{teaserfigure}

\maketitle
\pagestyle{plain}

\section{Introduction}
Wearable devices that enforce and augment our body are becoming pervasive nowadays. To augment human ability using robotic technologies~\cite{Sasaki17, xie19}, a wearable robotic limb is helpful in supporting the daily activities as a third arm. In this work, we focus on the physical augmentation with a compact robotic limb.

\begin{figure*}
    \centering
    \includegraphics[width=\textwidth]{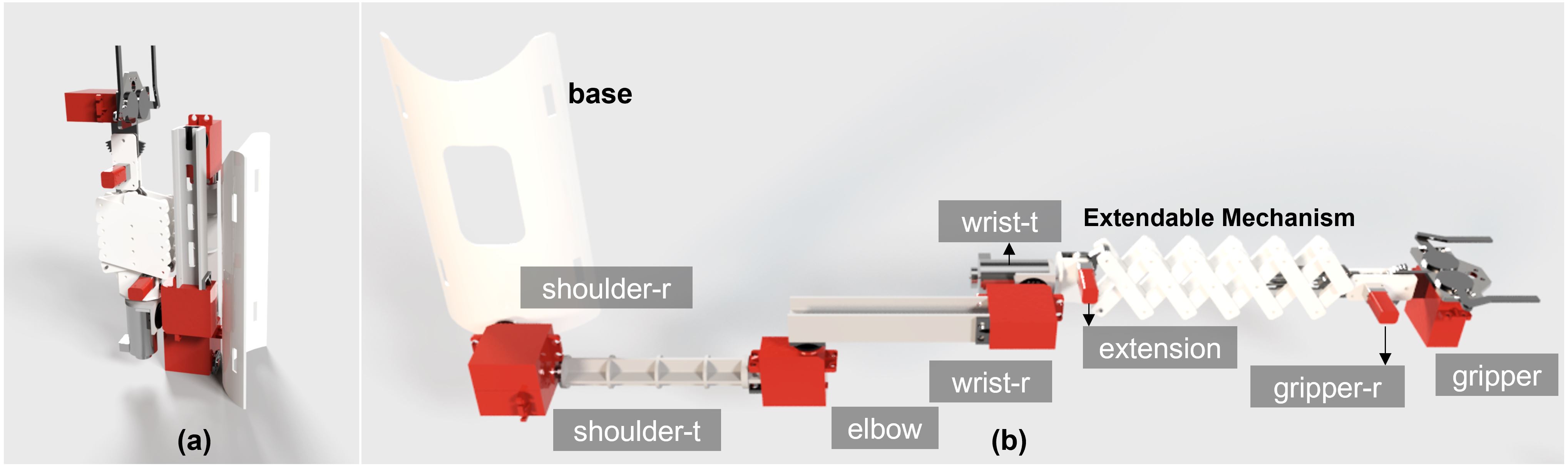}
    \caption{The proposed AugLimb device in compact state (a) and the expanded state (b). }
    \label{fig:system}
\end{figure*}

Previous works on robotic limbs are usually heavy and bulky, such as the supernumerary robotic limb for workers~\cite{Bright17}. The user may feel fatigue wearing these devices for long time usage, and the devices may cause occlusion to user's activities while not in usage. MetaLimbs proposed a pair of robotic arms carried on the user's back~\cite{Sasaki17}. However, this device may require large volume. xLimb adopted the storable and extendable mechanisms that reduce the burden exerted on the user~\cite{xlimb21}. In this work, we improve the controllability of robotic limb to larger degrees of freedom (DOFs) and the extendable mechanism with a double-layer scissor unit for stable and various system control.

In this work, we propose a compact robotic limb, AugLimb, that can transform from a compact state to expanded state with quite long achievable range by applying the extendable mechanism as shown in Figure~\ref{fig:teaser}. The main features of AugLimb are compact storage and extendable function. AugLimb is compact in idle status (Figure~\ref{fig:teaser}(a)) without interfere with the wearer's daily activities, and reach a far distance much longer than human arm (Figure~\ref{fig:teaser}(b)).

\section{Design and Implementation}

\subsubsection*{Device Design} Figure~\ref{fig:system} shows the proposed robotic limb of AugLimb in compact and expanded states. AugLimb has 7 DOFs and 1 extension unit. The whole joints include shoulder, elbow, wrist, and gripper parts. The shoulder and wrist parts have both rotational and twisting joints, shoulder-t and shoulder-r, wrist-t and wrist-r, respectively. The gripper part includes an unit of extendable mechanism with the extension joint, gripper and gripper-r joint for rotating the gripper. The supplementary video shows more details of the device design of AugLimb and its joint functions.  

\subsubsection*{Hardware Implementation} As shown in Figure~\ref{fig:system}(b), we used 5 servomotors and 4 direct current (DC) gear motors in our hardware design. In our prototype design, we adopted GX3370BLS motor for shoulder-r, DS3235SG motor for the shoulder-t, S53-20 motors for the elbow and wrist-r, SG92R motor for the gripper, a DC encoder gear motor (6V 10-1000rpm) with reduction gearbox (reduction ration 1:130) for the wrist-t, and GA12-N20 DC motors for the extension joints. Note that we used two DC motors for the extension joints to increase the torque force. For reduce the redundancy of robotic control, we did not implement the gripper-r joint in current prototype. We used Prusa I3 MK3S 3D printer to fabricate the arm base, shoulder, elbow, gripper and extension unit with PLA filament material. Without the control unit, the net weight of our implemented device is 640g. We used Arduino mega as control unit with 7 volts supply power. 

\subsubsection*{Extendable Mechanism} We adopt the double-layer scissor unit for the extendable mechanism to expand the achievable range of the gripper joint. With the proposed mechanism, the unit can be expanded to 250 mm, 3.6 times of the non-extension state (70 mm). For the whole robotic limb design, the maximum reachable length from the pivot of the base motor was 710 mm with the extended mode (630 mm without gripper). Therefore, AugLimb can achieve 2.5 times than the forearm length of a male adult (250 mm). Note that the gripper length is not included in measurement. 

\subsubsection*{Anticipated Applications} 
\begin{figure}
    \centering
    \includegraphics[width=\linewidth]{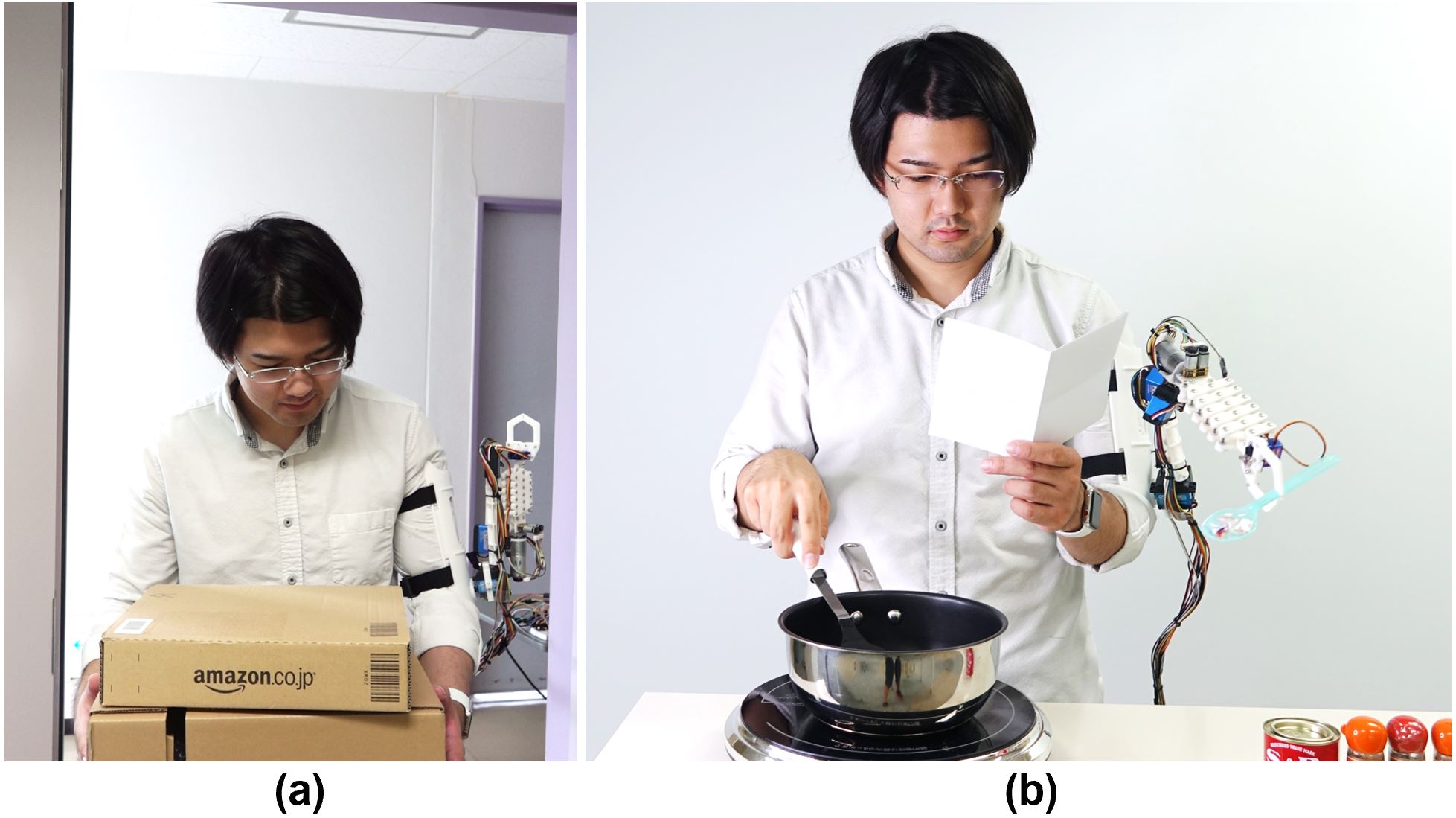}
    \caption{AugLimb can be transformed into compact state while the user walks through the narrow door (a), and can fetch the object far away in the cooking scenario (b).\vspace{-2.5mm} }
    \label{fig:application}
\end{figure}

Figure~\ref{fig:application} shows two examples of the anticipated applications with the proposed AugLimb device. Due to the advantages of compact size and extension mechanism, AugLimb could be useful for supporting various daily activities including walking, cooking, and working environments. 

\section{Discussion}

Automatic control of AugLimb from biological information is one issue to be solved in near future. This study verified the system design with the fabricated prototype which was controlled manually by the Wizard of Oz approach. It is difficult to apply foot movement tracking \cite{Sasaki17} for standing postures which are common in daily activities. The potential solutions may acquire motion data from the upper-body parts such as shoulder and wrist. The limited weight that AugLimb can lift can be improved by increasing the payload through the motors adoption, structure and material optimizations, such  as metal and carbon fiber materials. Please find more details of the proposed device in the introduction video\footnote{https://youtu.be/WyTDA7tjCjM}. 

\bibliographystyle{ACM-Reference-Format}
\bibliography{aug}

\end{document}